\documentclass[prl,twocolumn,showpacs,preprintnumbers,amsmath,amssymb]{revtex4}

\usepackage{graphicx}
\usepackage{dcolumn}
\usepackage{bm}
\usepackage{psfrag}
\usepackage{epsfig}
\usepackage{amsmath}
\usepackage{amssymb}
\usepackage{color}
\usepackage{mathbbol}

\newcommand{\proj}[1]{\ket{#1}\bra{#1}}
\newcommand{\bra}[1]{\langle #1 |}
\newcommand{\ket}[1]{| #1 \rangle}
\newcommand{\nn}{\nonumber}
\newcommand{\Bra}{\langle}
\newcommand{\Ket}{\rangle}

\newcommand{\id}{{\sf 1 \hspace{-0.3ex} \rule{0.1ex}{1.52ex}\rule[-.01ex]{0.3ex}{0.1ex}}}
\newcommand{\ignore}[1]{}

\begin{document}

\title{Density Matrix Renormalization Group in the Heisenberg Picture}

\author{Michael J. Hartmann$^{1,2,3}$}
\email{michael.hartmann@ph.tum.de}
\author{Javier Prior$^{2,3,4}$}
\author{Stephen R. Clark$^{4,5}$}
\author{Martin B. Plenio$^{2,3}$}
\affiliation{$^1\,$Technische Universit{\"a}t M{\"u}nchen, Physik Department I,
James Franck Str., 85748 Garching, Germany}
\affiliation{$^2\,$Institute for Mathematical Sciences, Imperial College London, SW7 2PG, United Kingdom}
\affiliation{$^3\,$QOLS, The Blackett Laboratory, Imperial College London, London SW7 2BW, United Kingdom}
\affiliation{$^4\,$Clarendon Laboratory, University of Oxford, Parks Road, Oxford OX1 3PU, United Kingdom}
\affiliation{$^5\,$Centre for Quantum Technologies, National University of Singapore, 3 Science Drive 2, Singapore 117543, Singapore}

\date{\today}

\begin{abstract}
In some cases the state of a quantum system with a large number of subsystems
can be approximated efficiently by the density matrix renormalization group,
which makes use of redundancies in the description of the state.
Here we show that the achievable efficiency can be much better when performing
density matrix renormalization group calculations in the Heisenberg picture,
as only the observable of interest but not the entire state is considered.
In some non-trivial cases, this approach can even be exact for finite bond dimensions.
\end{abstract}

\maketitle

%
\paragraph{Introduction --}
Quantum many-particle systems give rise to a number of 
intriguing phenomena such as quantum phase transitions, magnetic 
frustration, the existence of rare-earth magnetic insulators 
or high-temperature superconductivity. But as the size of the 
Hilbert space grows exponentially with the number of subsystems, 
the numerical simulation of such quantum many-body systems is 
difficult and often intractable.

In some cases, however, a quantum system does not explore its
entire Hilbert space and numerical approaches like the {\it 
density-matrix renormalization group} (DMRG) technique \cite{Wh91}, 
become efficient tools. DMRG can be understood as a 
variation over the set of {\it matrix product states} (MPS) \cite{FNW92,Rommer95}
whose size grows only polynomially with the number of subsystems.
Its success is linked to the existence of an upper bound for the 
entanglement of contiguous sub-blocks of the system under study 
\cite{Vi03,PZ07,ECP08}. This approach is therefore expected to 
work particularly well for the ground state of one-dimensional 
gapped systems, in which correlation functions decay exponentially 
and the entanglement entropy saturates, satisfying an ``area law'' 
\cite{Area,ECP08}.
There are of course situations in which no upper bound to the 
entanglement in the system exists or where it grows in time. In
such cases the performance of DMRG deteriorates. This is typically 
the case for the dynamics of non-equilibrium states, as exemplified 
in recent studies of sudden quenches to Bose-Hubbard Hamiltonians 
\cite{CDEO07}. Due to its dynamical production, the entanglement per
unit area may grow linearly in time in those scenarios \cite{Calabrese,SchuchWVC08}.
To achieve a fixed precision DMRG algorithms hence need to use matrix dimensions 
that grow exponentially in time rendering them inefficient 
\cite{SchuchWVC08}.

An increasing number of experimental settings, including arrays of 
Josephson junctions \cite{FZ01}, ultra cold atoms in optical lattices 
\cite{BDZ07}, ion traps \cite{ion} and arrays 
of coupled micro-cavities \cite{HBP06}, offer the possibility
to generate effective many-particle systems. Hence, dynamical studies of 
quantum many-particle systems are expected to receive increasing 
attention in the future. Moreover in real experimental situations,
such systems will typically suffer from decoherence and dissipation and
hence evolve into mixed states whose numerical 
description is even more demanding. It is therefore desirable to
develop new more efficient methods for such problems or alternatively
to improve existing DMRG methods further.

In this letter we describe an approach to enhance the performance 
of DMRG in time-dependent settings by computing directly the observable 
of interest and propagating its evolution in the Heisenberg picture. Our
approach avoids calculating components of the quantum states that are irrelevant 
to the observable of interest and hence contrasts the standard DMRG approach,
which generates a quantum state for the entire quantum system in the Schr\"odinger
picture.  
DMRG performed in the Heisenberg and Schr\"odinger picture are thus
complementary. Whereas Schr\"odinger picture simulations calculate the entire
state and subsequently allow to compute any observable of interest,
Heisenberg picture calculations only consider one operator whose expectation
value can in turn be computed for any initial state.

In the sequel, we demonstrate that DMRG performed in the Heisenberg picture 
(H-DMRG) can have significant advantages for numerical simulations 
of quantum many-particle dynamics. These advantages become most 
significant in open system dynamics described by mixed states but 
can also be demonstrated rigorously for certain exactly solvable 
systems. We find numerical indications for a saturation of the 
block entanglement in the Heisenberg picture for increasing system 
size which suggest that H-DMRG has superior efficiency in many cases.

\paragraph{ Main part --}

For linear chains of interacting subsystems,
we consider the evolution of operators such as
${\cal X}_m(t) = U(t)\left( \id^{\otimes m-1}\otimes {\cal X} \otimes
\id^{\otimes N-m} \right) U(t)^{\dagger}$,
where ${\cal X}$ is a Hermitian operator acting on site $m$,
and use a matrix-product representation
\begin{equation} \label{MPOrep}
  {\cal X}_m = \sum_{i_1,\ldots, i_N = 0}^3 \text{tr}\bigl[A_{i_1}^{(1)}
  \ldots A_{i_N}^{(N)} \bigr] P_{i_1}\otimes \ldots \otimes P_{i_N}
\end{equation}
with suitable $d \times d$-dimensional matrices $A_{i_l}^{(l)}$ and
the canonical operator basis $\{P_0,P_1,P_2,P_3\}$ with $(P_m)_{i,j}
=\delta_{m,2i+j}$ for $i,j\in\{0,1\}$.
We focus our study on dynamics of the anisotropic Heisenberg
Hamiltonian for a chain of $N$ spins,
\begin{equation} \label{HeisenH}
 H = \sum_{j=1}^N B_z \sigma_j^z + \sum_{j=1}^{N-1}
 \sum_{\alpha=x,y,z} J_{\alpha} \sigma_j^{\alpha} \sigma_{j+1}^{\alpha}\, ,
\end{equation}
as this model is known to exhibit dynamics that is numerically hard to simulate.
In eq. (\ref{HeisenH}), $B_z$ is an applied magnetic field, $J_x, J_y$ and $J_z$
are spin-spin couplings and $\sigma_j^x$, $\sigma_j^y$ and $\sigma_j^z$ the Pauli
operators at site $j$.

\paragraph{Exact results --}

It is noteworthy that the time evolution of certain operators
can actually be represented {\em exactly} by a matrix product 
operator with fixed finite bond dimension. For Hamiltonians 
of the form of eq. (\ref{HeisenH}) with $J_z = 0$,
all local operators that transform under the Jordan-Wigner 
transformation \cite{LSM61} into local fermionic operators
remain exact matrix product operators with fixed finite dimension
for all times. Examples of such operators are $\sigma_m^z$ whose
time evolution is an exact matrix product operator for matrix
dimension $d=4$ and generally any product of Pauli-operators
with an even number of $\sigma^x$ or $\sigma^y$ operators and 
any number of $\sigma^z$ operators.

To see this, let us first define the fermionic
annihilation and creation operators \cite{LSM61},
$c_m = \prod_{j=1}^{m-1} \sigma^{z}_j \, (\sigma^{x}_m + i \sigma^{y}_m)/2$.
In terms of $c_m$ and $c_m^{\dagger}$, the Hamiltonian (\ref{HeisenH})
with $J_z = 0$ reads
$H = - B \sum_{j=1}^N (2 c^{\dagger}_j c_j - 1)
+ J_x \sum_{j=1}^{N-1} (c_j^{\dagger} - c_j)(c_{j+1}^{\dagger} + c_{j+1})
- J_y \sum_{j=1}^{N-1} (c_j^{\dagger} + c_j)(c_{j+1}^{\dagger} - c_{j+1}).$
Given that the Hamiltonian is quadratic in $c_m$ and $c_m^{\dagger}$, 
the Heisenberg time evolution of an individual Heisenberg operator 
such as $c_m (t)$ is found to be
$c_m (t) = \sum_{j=1}^{N} \left( \alpha_j(t) c_j + \beta_j(t) c_j^{\dagger} \right)$.
%
In the fermionic picture this may be written as matrix product operator
with matrices of dimension $2$ as it is essentially the same as a W-state.
Rewriting the rhs in terms of Pauli operators we find
$c_m (t) =
 \sum_{j=1}^{N} \left( \alpha_j(t) \prod_{l=1}^{j-1} \sigma^{z}_l \sigma^+_j
 + \beta_j(t) \prod_{l=1}^{j-1} \sigma^{z}_l \sigma^-_j \right)$,
%
where $\sigma^{\pm} = \frac{1}{2} \left( \sigma^{x} \pm i \sigma^{y} 
\right)$. This in turn may be written as a matrix product operator 
of the form eq. (\ref{MPOrep}) whose matrices have the structure
$A_0^{(1)} = P_1, \:
 A_0^{(m)} = P_0 + P_3, \:
 A_0^{(N)} = P_0, \:
 A_1^{(1)} = \alpha_1 (t) P_0, \:
 A_1^{(m)} = \alpha_m (t) P_2, \:
 A_1^{(N)} = \alpha_N (t) P_2, \:
 A_2^{(1)} = \beta_1 (t) P_0, \:
 A_2^{(m)} = \beta_m (t) P_2, \:
 A_2^{(N)} = \beta_N (t) P_2, \:
 A_3^{(1)} = - P_1, \:
 A_3^{(m)} = P_0 - P_3, \:
 A_3^{(N)} = P_0$.
 
As every spin operator may be expressed as a sum of products
of fermionic operators we can now understand the above 
observations. For example, because $\sigma_m^z=2 c^{\dagger}_m c_m -1$
we can write it as a product of two matrix product operators
each with dimension $2$, so that $\sigma_{k}^{z}$ is a matrix
product operator with dimension at most $4$.
This reasoning also holds for models with disorder,
i.e. where the magnetic field or the couplings depend on the lattice
site ($B_z(j), J_x(j)$ and $J_y(j)$), which can not be diagonalized
via Fourier and Bogolubov transformations \cite{LSM61}.  Analogous 
conclusions hold for quasi-free bosonic systems.

This observation demonstrates that a DMRG simulation in the 
Heisenberg picture may be considerably more efficient, even 
exact, in cases where the same approach in the Schr{\"o}dinger
picture is provably inefficient \cite{SchuchWVC08}.
In contrast to the Schr{\"o}dinger picture, the block entanglement
in the Heisenberg picture (considering the four operators 
$P_0,P_1,P_2,P_3$ as basis-vectors of a 4-dim Hilbert 
space for each site) is bounded for all times. This difference
in entanglement scaling in the two pictures obviously can not
hold for all settings
\footnote{It is clear that there must be 
situations in which also the simulation in the Heisenberg 
picture must fail. If this was not the case, it would be 
possible to simulate efficiently quantum algorithms as 
their final readout consists of single spin measurements 
only.}. Nonetheless we find indications for a saturation
in the scaling of block entanglement in numerical simulations
for more general models.

\paragraph{Numerical results --}

We now turn to compare the numerical efficiency of H-DMRG with
that of DMRG in the Schr\"odinger picture \cite{Wh91,FNW92,Rommer95,Vi03}.
For the dynamics of
pure states, we have seen that there are examples where DMRG becomes 
exact thanks to a very favorable behavior of entanglement. In 
general one expects the use of H-DMRG to be advantageous only 
where the entanglement scaling for the state is drastically worse
than for the operator to be evolved. This is due to the following
reason: If a quantum state has the matrix product representation
$\ket{\Psi} = \sum_{i_1\ldots i_N} \text{tr}\bigl[A_{i_1}^{(1)}
\ldots A_{i_N}^{(N)} \bigr]  \ket{i_1} \otimes \ldots \otimes \ket{i_N}$
with matrix dimension $d$, then the operator $\proj{\Psi}$ has the
matrix product representation
\begin{equation} \label{MPOrep2}
   \proj{\Psi} = \sum_{i_1, \ldots, i_N = 0}^3 \text{tr}\bigl[B_{i_1}^{(1)}
  \ldots B_{i_N}^{(N)} \bigr]  \proj{i_1} \otimes \ldots \otimes \proj{i_N},
\end{equation}
where $B_{i_l}^{(l)} = A_{i_l}^{(l)} \otimes \left( A_{i_l}^{(l)} \right)^{\dagger}$
and hence the $B$ matrices have dimension $d^2$.
The matrix product representation of an operator is thus expected to
require matrix dimension $d^2$ in situations, where the representation of
a state only requires $d$ and is therefore much more efficient.

The situation is different if decoherence and dissipation is present 
as then the evolution of an operator must be considered in both,
Heisenberg and Schr\"odinger picture. Dissipation may be described 
by local Lindblad terms leading to a master equation for the dynamics 
of the density matrix $\varrho$ of the form \cite{GZbook},
\begin{eqnarray} \label{mastereq}
 \dot{\varrho} & = & -i \left[ H, \varrho \right]
+ \sum_{j=1}^N \frac{\Gamma_d}{2} \left(2 \sigma_j^- \varrho \sigma_j^+ - \sigma_j^+ \sigma_j^- \varrho - \varrho \sigma_j^+ \sigma_j^- \right)\nn \\
& + & \sum_{j=1}^N \frac{\Gamma_u}{2} \left(2 \sigma_j^+ \varrho \sigma_j^- - \sigma_j^- \sigma_j^+ \varrho - \varrho \sigma_j^- \sigma_j^+ \right),
\end{eqnarray}
where $\Gamma_d$ and $\Gamma_u$ are the respective damping rates.
When the description is transferred into the Heisenberg picture, the same
dynamics is described by the equation,
\begin{eqnarray} \label{heisenmaster}
 \dot{{\cal X}} & = & i \left[ H, {\cal X} \right]
+ \sum_{j=1}^N \frac{\Gamma_d}{2} \left(2 \sigma_j^+ {\cal X} \sigma_j^- - \sigma_j^+ \sigma_j^- {\cal X} - {\cal X} \sigma_j^+ \sigma_j^- \right)\nn \\
& + & \sum_{j=1}^N \frac{\Gamma_u}{2} \left(2 \sigma_j^- {\cal X} \sigma_j^+ - \sigma_j^- \sigma_j^+ {\cal X} - {\cal X} \sigma_j^- \sigma_j^+ \right),
\end{eqnarray}
for a Heisenberg picture operator ${\cal X}(t)$. In the following 
we compare the results of numerical simulations in the Schr{\"o}dinger 
(eq. \ref{mastereq}) \cite{ZV04} and Heisenberg picture (eq. \ref{heisenmaster}).

\begin{figure}
\psfrag{A}{\hspace{-1cm} \bf a}
\psfrag{B}{\hspace{-1.2cm} \bf b}
\psfrag{C}{\hspace{-1.2cm} \bf c}
\includegraphics[width=\linewidth]{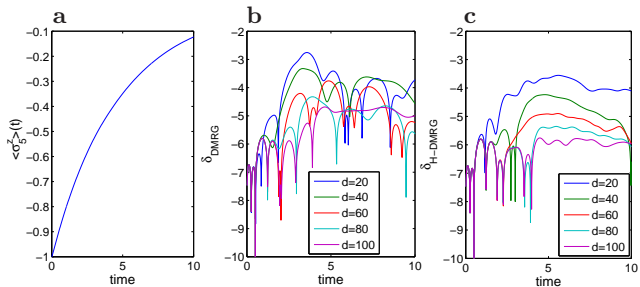}
\caption{\label{N10_1} The time evolution, $\Bra \sigma_5^z \Ket (t)$ for
a model described by eqs. (\ref{mastereq}) respectively (\ref{heisenmaster}) with parameters
$N = 10$, $B_z = 0.8$, $J_x = 0.5$, $J_y = 0.4$, $J_z = 0.01$, $\Gamma_u = 0.1$
and $\Gamma_d = 0.1$.
{\bf a}: The exact solution,
{\bf b}: $\delta_{\text{DMRG}}$ for DMRG-simulations in the Schr\"odinger picture
for $d=20$ (blue), $d=40$ (green), $d=60$ (red), $d=80$ (cyan) and $d=100$ (magenta),
{\bf c}: $\delta_{\text{H-DMRG}}$ for DMRG-simulations in the Heisenberg picture.}
\end{figure}

In our first example we choose the parameters of the model to be,
$N = 10$, $B_z = 0.8$, $J_x = 0.5$, $J_y = 0.4$, $J_z = 0.01$,
$\Gamma_u = 0.1$ and $\Gamma_d = 0.1$ to allow for comparison with
exact results.
We simulate the time evolution of the operator $\sigma_5^z (t)$,
where the initial state is all spins pointing down in $z$-direction,
$\ket{\phi_0} = \ket{\downarrow, \dots, \downarrow}$
($\sigma^z \ket{\downarrow} = - \ket{\downarrow}$).
Figure \ref{N10_1}{\bf a} shows the exact solution, that is a 2nd order
Runge-Kutta integration with time steps $dt = 0.01$ of eq. (\ref{mastereq}).
All our DMRG simulations also use 2nd order integrations with $dt = 0.01$.
The errors of DMRG-simulations in the Schr\"odinger picture,
$\delta_{\text{DMRG}} = \log_{10}\left(|\Bra \sigma_5^z \Ket_{\text{exact}} (t)
- \Bra \sigma_5^z \Ket_{\text{DMRG}} (t)|\right)$, and the Heisenberg picture,
$\delta_{\text{H-DMRG}} = \log_{10}\left(|\Bra \sigma_5^z \Ket_{\text{exact}} (t)
- \Bra \sigma_5^z \Ket_{\text{H-DMRG}} (t)|\right)$, are shown in
figures \ref{N10_1}{\bf b} and {\bf c} respectively
(Calculations with time steps $dt = 0.001$ produced indistinguishable results).
The Heisenberg picture simulations show a significantly higher accuracy than the
Schr\"odinger picture results. Moreover the accuracy improvement with
increasing matrix dimension $d$ is more pronounced in the Heisenberg picture,
suggesting an unfavorable scaling of entanglement in the Schr{\"o}dinger picture.

\begin{figure}
\psfrag{A}{\hspace{-1.6cm} \bf a}
\psfrag{B}{\hspace{-1.6cm} \bf b}
\psfrag{C}{\hspace{-1.6cm} \bf c}
\psfrag{D}{\hspace{-1.6cm} \bf d}
\includegraphics[width=0.49\linewidth]{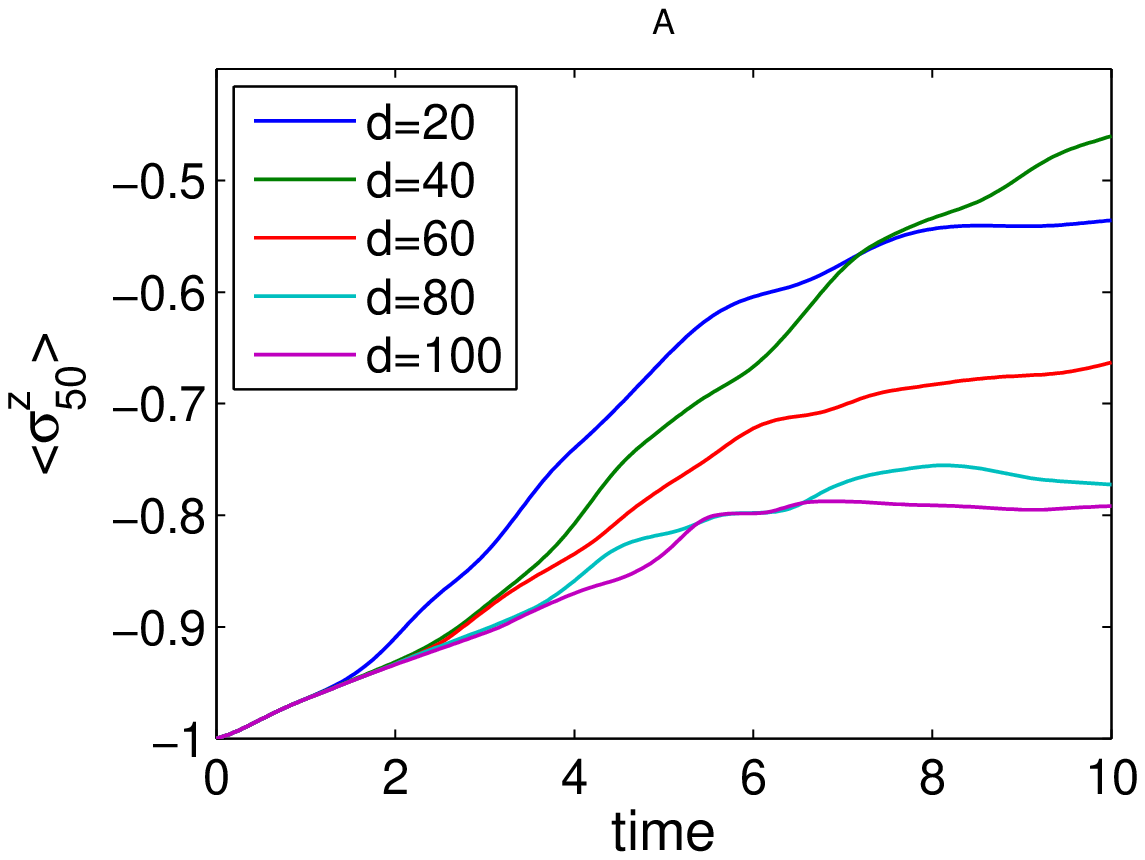}
\includegraphics[width=0.49\linewidth]{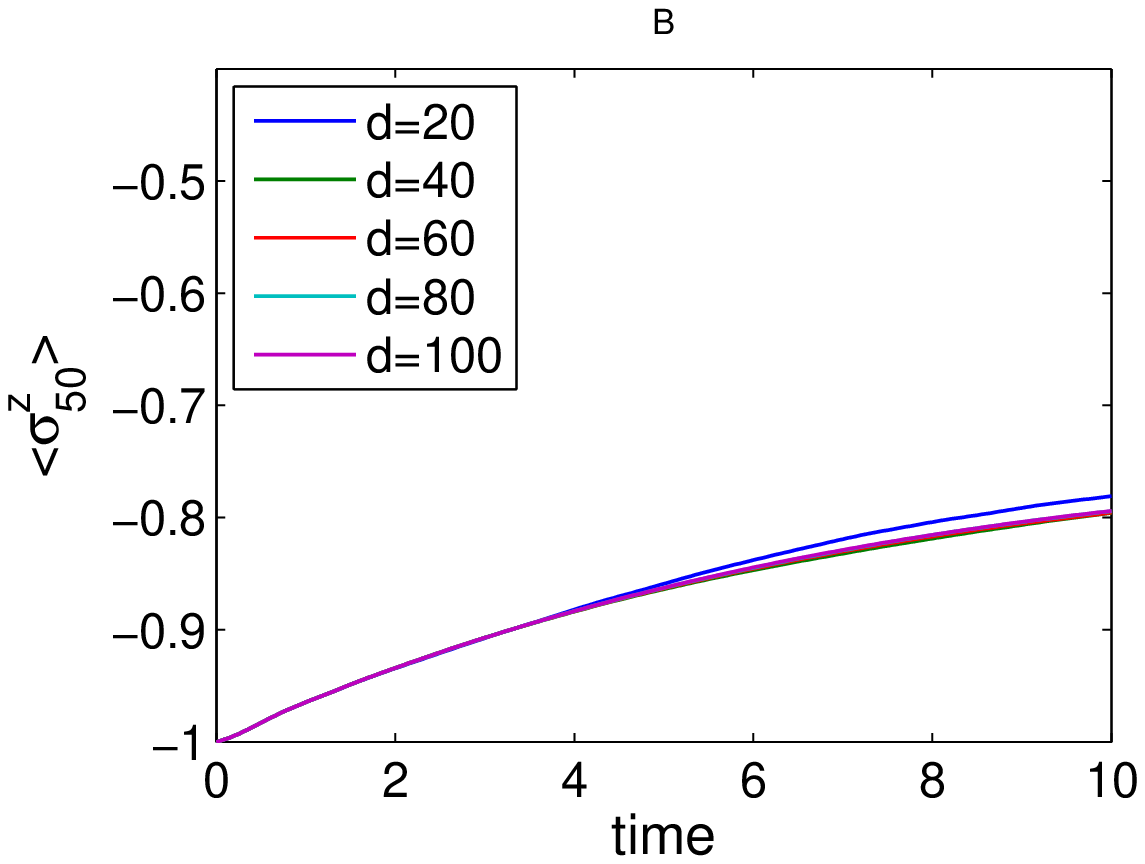}\\
\includegraphics[width=0.49\linewidth]{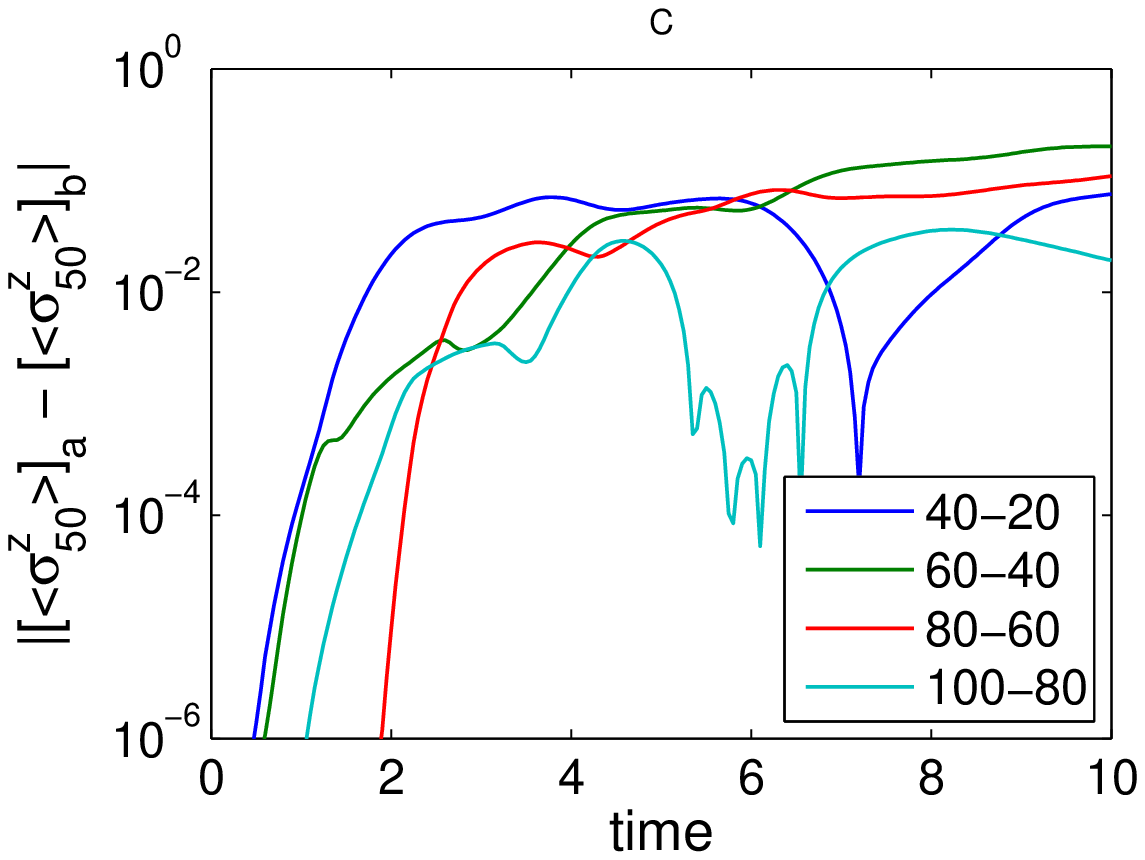}
\includegraphics[width=0.49\linewidth]{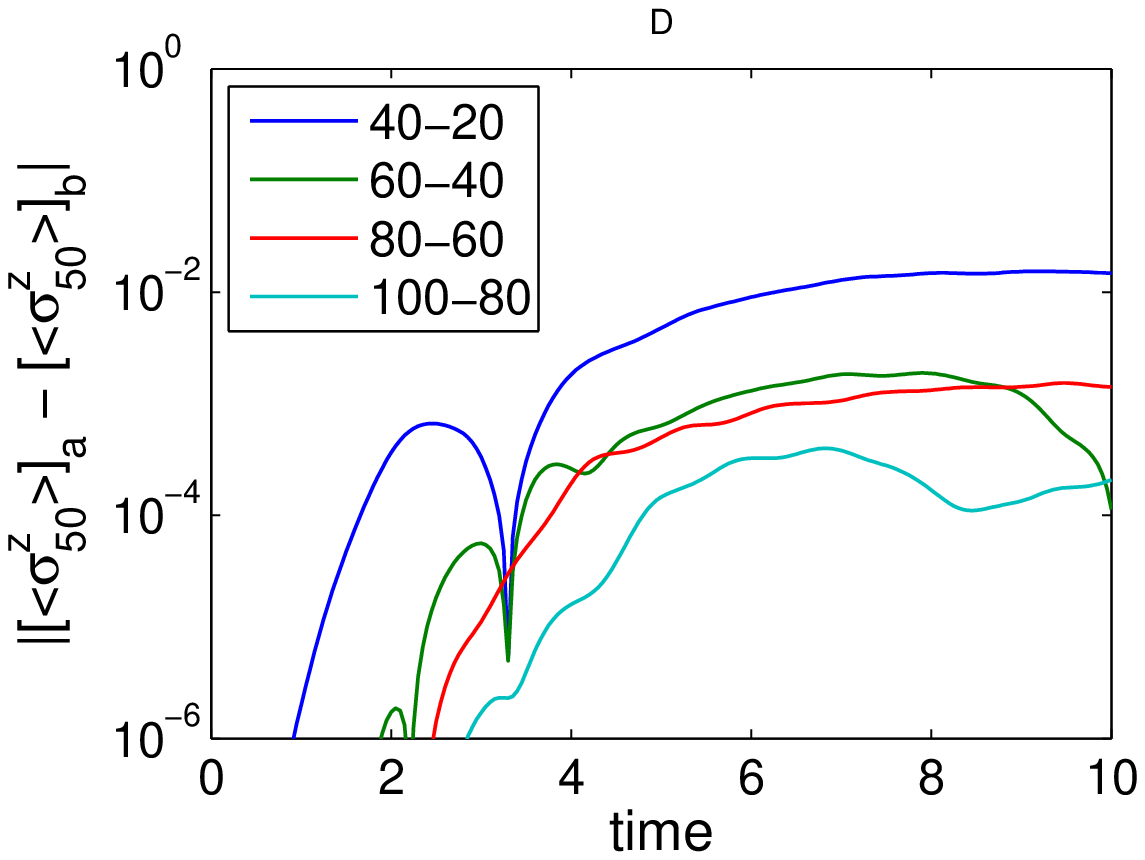}
\caption{\label{N100_1} Dynamics of a chain with $N=100$ spins and
$B_z = 0.8$, $J_x = 0.5$, $J_y = 0.4$, $J_z = 0.01$, $\Gamma_u = 0.01$
and $\Gamma_d = 0.1$.
{\bf a:} $\Bra \sigma_{50}^z \Ket (t)$ as given by eq. (\ref{mastereq}) (DMRG)
for $d=20$ (blue), $d=40$ (green), $d=60$ (red), $d=80$ (cyan) and $d=100$ (magenta).
{\bf b:} $\Bra \sigma_{50}^z \Ket (t)$ as given by eq. (\ref{heisenmaster}) (H-DMRG)
for $d=20,40,60,80$ and $100$.
{\bf c:} $\left| \left[ \Bra \sigma_{50}^z \Ket (t) \right]_{d1}
- \left[ \Bra \sigma_{50}^z \Ket (t) \right]_{d2} \right|$ as given by eq. (\ref{mastereq}) (DMRG)
for $(d1,d2) = (40,20)$ (blue), $(d1,d2) = (60,40)$ (green), $(d1,d2) = (80,60)$ (red)
and $(d1,d2) = (100,80)$ (cyan).
{\bf d:} $\left| \left[ \Bra \sigma_{50}^z \Ket (t) \right]_{d1}
- \left[ \Bra \sigma_{50}^z \Ket (t) \right]_{d2} \right|$ as given by
eq. (\ref{heisenmaster}) (H-DMRG) for $(d1,d2) = (40,20), (60,40), (80,60)$ and $(100,80)$.
With increasing bond dimension, results converge much faster in the Heisenberg 
than in the Schr\"odinger picture.}
\end{figure}

In a second example we consider a chain of $N=100$ spins, with slightly different
parameters to show the generality of our findings, and compare DMRG results
in the Schr\"odinger and Heisenberg picture. Here, $B_z = 0.8$, $J_x = 0.5$, $J_y = 0.4$,
$J_z = 0.01$, $\Gamma_u = 0.01$ and $\Gamma_d = 0.1$.
Figures \ref{N100_1}{\bf a} and \ref{N100_1}{\bf b }show $\Bra \sigma_{50}^z \Ket (t)$
as calculated in the Schr\"odinger (\ref{N100_1}{\bf a}) and
Heisenberg picture (\ref{N100_1}{\bf b}).
Since it is not possible to compare these values to exact results for $N=100$,
we test the convergence of the obtained results with increasing matrix dimension, $d$.
This convergence is shown in figure \ref{N100_1}{\bf c}
for the Schr\"odinger and in figure \ref{N100_1}{\bf d}
for the Heisenberg picture, where we plotted the differences between results that were
obtained with different matrix dimensions $d1$ and $d2$,
$\left| \left[ \Bra \sigma_{50}^z \Ket (t) \right]_{d1} - \left[ \Bra \sigma_{50}^z \Ket (t) \right]_{d2} \right|$.
The convergence is found to be much faster in the Heisenberg than in the Schr\"odinger picture.
We found the advantage of the Heisenberg picture to be of a comparable significance as
in figure \ref{N100_1} already for a chain with the same parameters but $N = 40$ spins.

In our simulations we use the {\it time evolved block decimation} (TEBD) algorithm
introduced in \cite{Vi03}. This algorithm, at each step, truncates 
the reduced density matrices of all considered bipartitions by only 
keeping the states corresponding to their $d$ largest eigenvalues
(Here $d$ is the dimension of the employed matrices.).
We compute the truncation error by summing up all truncated eigenvalues
of the reduced density matrices \cite{Vi03} at each time step. The truncation
errors at each time step, $\epsilon_{t}$, are then cumulatively summed up. The resulting
quantity, $\epsilon = \sum_{t} \epsilon_{t}$, is thus an upper bound to approximation errors
due to matrix truncation.

To enable a comparison of the accuracies of the matrix truncations,
we set Heisenberg and Schr\"odinger representations on the same footing
by normalizing the representations of eqs. (\ref{MPOrep2}) 
and (\ref{MPOrep}) in the Frobenius norm, i.e. for a $m\times m$ 
matrix $X$ we set $\sum_{i,j=1}^{m} |X_{i,j}|^2 = 1$. With this normalization,
the matrix representations of $\varrho$ and ${\cal X}$ both have the structure of
Matrix Product States where the eigenvalues of reduced matrices sum up to unity
and can be interpreted as Schmidt coefficients\footnote{Since this normalization
is not preserved by eqs. (\ref{mastereq}) and (\ref{heisenmaster}),
the representations need to be renormalized after each time step.}.

Since we compare truncation errors in two different representations 
it is not obvious that lower truncation errors in one representation 
imply a better approximation for the expectation value of an observable 
or vice versa. Indeed, for the short chains used in the example in 
figure \ref{N10_1} we found comparable truncation errors in both 
approaches even though the error in the relevant observable is 
much smaller when using H-DMRG\footnote{Obviously, the calculated
expectation values need to be multiplied by the inverse norms to obtain
the physically relevant results.}.

\begin{figure}
  \includegraphics[width=0.6\linewidth]{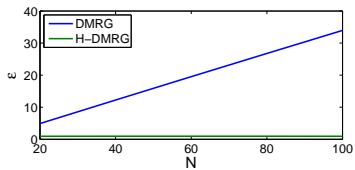}
  \caption{Truncation errors $\epsilon$ at $t = 10$ for Schr\"odinger picture DMRG (blue) and
  H-DMRG (green) for $N = 20, 40, 60$ and $80$ and fixed bond dimension $d=60$. 
  $B_z = 0.8$, $J_x = -0.5$,
  $J_y = 0.4$, $J_z = 0.1$, $\Gamma_u = 0.1$ and $\Gamma_d = 0.2$.
  H-DMRG truncations saturate at $\epsilon \le 0.96$.}
  \label{truncations}
\end{figure}

On the other hand, the truncation errors appear to be significantly 
lower in H-DMRG for longer chains. We have therefore investigated 
the scaling of the truncation error with the system size for fixed bond
dimension $d$. Importantly, we observe
that in contrast to Schr\"odinger picture DMRG, the truncation 
errors $\epsilon$ are found to saturate in H-DMRG for fixed bond
dimension if the system size is increased. Figure \ref{truncations} 
shows the truncation errors $\epsilon$ at $t = 10$ for Schr\"odinger picture 
DMRG (blue) and H-DMRG (green) for $d = 60$ and 
chain length $N = 20, 40, 60, 80$ and $100$. For the remaining parameters we
chose a set, that is again different from those in figures \ref{N10_1} and \ref{N100_1} 
with $B_z = 0.8$, $J_x = -0.5$, $J_y = 0.4$, $J_z = 0.1$, $\Gamma_u = 0.1$ 
and $\Gamma_d = 0.2$.

This clear difference in the scaling further 
corroborates the idea that the Heisenberg and Schr{\"o}dinger 
picture are qualitatively different in regards to their entanglement 
scaling even beyond the exactly solvable models discussed earlier. 
In fact, this feature hints at a saturation of the entanglement
of bipartitions in the Heisenberg picture.
It will be an interesting challenge for future work to provide 
analytical arguments to support the numerical findings presented 
here and to demonstrate more rigorously the superior efficiency
of H-DMRG when applied to mixed state evolutions beyond the 
numerical findings here.

{\em Acknowledgements --}
This work is part of EU Integrated Project QAP (contract 015848),
the EPSRC QIP-IRC (GR/S82176/0) and the DFG Emmy Noether project
HA 5593/1-1. It was supported by EPSRC grant 
EP/E058256, the Fundaci\'on S\'eneca grant 05570/PD/07 and the Royal Society.

\end{document}